# Toward more performant eye safe lasers: effect of increasing sensitizer amount in $Yb^{3+}$,$Er^{3+}$:YAG transparent ceramic on its spectral characteristics.

Agnieszka Szysiak[1*], Robert Tomala[2,3], Helena Węglarz[1], Juraj Kajan[4,5], Mirosław Słobodzian[1], Mykhailo Chaika[1,2,3*]

[1]Łukasiewicz Research Network - Institute of Microelectronics and Photonics, Aleja Lotników 32/46, Warsaw 02-668, Poland.

[2]Graphene Energy Ltd, Curie-Sklodowskiej Str. 55/61, 50-369 Wroclaw, Poland

[3]Institute of Low Temperature and Structure Research, Polish Academy of Sciences, Okólna 2, 50 – 422 Wrocław, Poland

[4]Institute of Competitiveness and Innovations, University of Žilina, Univerzitná 1, Zilina, 010 26 Slovakia

[5]Advanced Technology Crystals s.r.o., Legionárska 1, Zilina, 010 26 Slovakia

**Abstract**

Developing efficient $Er^{3+}$,$Yb^{3+}$:YAG eye-safe lasers is a priority of modern laser technology. This paper focuses on the influence of the concentration of $Yb^{3+}$ ions on the spectroscopic properties of $Er^{3+}$,$Yb^{3+}$:YAG transparent ceramics. Four samples with different concentrations of $Yb^{3+}$ ions were prepared by solid-state reaction sintering. The study revealed the influence of $Yb^{3+}$ ions on the microstructure and the sintering process. A high concentration of $Yb^{3+}$ ions leads to the formation of $Y^{3+}$-rich impurity phases and causes segregation of $Er^{3+}$ and $Yb^{3+}$ and $Si^{4+}$ into these phases. The influence of $Yb^{3+}$ ions on the shape of emission spectra and the lifetimes of both $Er^{3+}$ and $Yb^{3+}$ ions was shown. Changes in the spectroscopic properties were ascribed to increase in neat transfer between $Yb^{3+}$ and $Er^{3+}$ ions. IQE of $Er^{3+}$ and $Yb^{3+}$ luminescence were calculated, and optimal $Er^{3+}$/$Yb^{3+}$ ions ratio were proposed.

\* agnieszka.Szysiak@imif.lukasiewicz.gov.pl

\* m.chaika@grapheneenergy.pl

# 1. Introduction

The recent progress in the laser technology makes it possible to achieve good performance for laser sources with various wavelengths from infrared (IR) to visible (Vis) and ultraviolet (UV) [1–3]. Therefore, the lasers are increasingly used in the everyday life. But the laser radiation can be dangerous to human health, especially when directed towards an eye. 1.5 μm lasers are considered eye-safe ones. Wavelengths in this region are absorbed in anterior portions of the eye (cornea) and therefore never reach the retina [4,5]. Despite its importance, the efficiency of 1.5 μm lasers is far from optimal [6]. Development of the efficient eye-safe lasers is one of the priorities of modern laser technologies.

The properties of lasers mostly depend on the type of active element [7]. Active elements for 1.5 μm lasers are mostly based on $Er^{3+}$ - doped materials. The disadvantage of $Er^{3+}$ ions is small absorption cross sections which makes it inefficient for direct pumping. Due to efficient energy transfer between $Er^{3+}$ ions [8], an increase in the concentration of $Er^{3+}$ ions causes an increase in the efficiency of the nonradiative recombination processes leading to decline in the efficiency of $Er^{3+}$ emission for highly doped materials. One of the possible solutions is co-doping with $Yb^{3+}$ ions, where $Yb^{3+}$-$Er^{3+}$ ions function as a donor-acceptor pair [9]. The main advantage of $Yb^{3+}$ ions is large absorption cross-section and efficient energy transfer to other ions [10]. Therefore, one of the most efficient 1.5 μm lasers is based on $Er^{3+},Yb^{3+}$:YAG ($Er_{3x}Yb_{3y}Y_{(1-x-y)3}Al_5O_{12}$). Despite high number of papers regarding the properties of such materials, their laser efficiency is far from perfect.

The increase in the lasing efficiency of the $Er^{3+},Yb^{3+}$:YAG materials is possible by increasing the efficiency of energy transfer from $Yb^{3+}$ ions to $Er^{3+}$ ones. The efficiency of energy transfer depends on a number of parameters, the most important of which is the distance between energy donors and acceptors [11,12]. The decrease in the distance increases the efficiency of energy transfer according to inverse sixth-power dependence [13]. The distance between the $Yb^{3+}$ and $Er^{3+}$ ions can be modified through the change in the concentrations of these ions [14]. The increase in the concentration of these ions caused an increase in the efficiency of energy transfer improving the laser properties [15]. However, together with an increase in the energy transfer between $Yb^{3+}$ and $Er^{3+}$ ions, the increase in their concentrations increases the probability of non-radiative relaxation processes, which negatively affects the properties of such materials. Therefore, the development of efficient $Er^{3+},Yb^{3+}$:YAG materials is possible by tuning the concentrations of both $Er^{3+}$ and $Yb^{3+}$ ions.

The present paper focuses on the study of the influence of $Yb^{3+}$ concentration on the spectroscopic properties of $Er^{3+},Yb^{3+}$:YAG transparent ceramics.

## 2. Experimental

The $Er^{3+},Yb^{3+}$:YAG transparent ceramics were obtained by Solid-State Reaction Sintering (SRS) in vacuum [16]. High purity $Al_2O_3$ (purity >99.99%, Taimei Chemicals), $Y_2O_3$ (purity >99.90%, Nippon Yttrium), $Er_2O_3$ (purity >99.999%, Metall Rare Earth), and $Yb_2O_3$ (purity >99.99%, Metall Rare Earth) and Duramax B-1000 (The Dow Chemical) as a binder, Dolapix CE 64 (Zschimmer & Schwarz) as a dispersing agent and laboratory prepared $[SiO_{2.5}N(CH_3)_4]_8 \cdot \sim 60 \cdot H_2O$ sintering aid (further called TAOS) were used as starting materials. TAOS was utilized in the synthesis, which has a freeze-granulation step. It was prepared using tetraethylammonium hydroxide (TMAH), methanol (MeOH), tetraethyl orthosilicate (TEOS), and water($H_2O$). These components were mixed, evaporated, dried, and dissolved to form a 10% aqueous solution. Powders were taken in stoichiometric ratio, concentrations of $Er^{3+}$ and $Yb^{3+}$ were taken in order to replace $Y^{3+}$ ions. $Er_2O_3$ and $Yb_2O_3$ powders were weighted precisely to obtain erbium, and ytterbium content of 0.5 at.% and 0-20 at.%, respectively. Homogenization was performed by ball milling for 2 hours using high purity $Si_3N_4$ balls. The milled slurry was sprayed into liquid nitrogen to form granules and afterwards freeze dried for 20 hours in a freeze-dryer. The obtained granulates were uniaxially pressed into disk-shaped samples 20 mm in diameter and further densified using a cold isostatic pressing at 120 MPa. The samples were calcined in air at 950°C to remove organic additives. The sintering was performed at 1750°C for 6 hours in vacuum furnace. In order to reduce oxygen vacancy concentration, the samples were annealed in air at 1600°C for 2 hours [17–20]. As a result, the concentration series of $Er^{3+},Yb^{3+}$:YAG transparent ceramics were obtained (Fig. S1). The synthesized samples and dopant concentrations are listed in the Table 1.

Special attention should be paid to the fact that the sample labeled as Er(0.5)Yb(<0.01) was synthesized as a Yb – free sample. However, luminescence measurements revealed the presence of small concentrations of $Yb^{3+}$ ions. The excitation laser power used for the Er(0.5)Yb(<0.01) sample was two orders of magnitude higher, while the luminescence intensity was two order of magnitude lover than that of the other samples. This indicates a low $Yb^{3+}$ concentration in the Er(0.5)Yb(<0.01) sample. The accuracy of the measured absorption spectra allow for the clear detection of $Yb^{3+}$ ions at 0.01 at.% in the samples. No sign of $Yb^{3+}$ absorption peaks were found in the Er(0.5)Yb(<0.01) samples, indicating that the actual concentration of $Yb^{3+}$ ions is lower than 0.01 at.%.

Table 1: The list of synthesized $Er^{3+}$,$Yb^{3+}$:YAG transparent ceramics where T is the temperature and t is the time of vacuum sintering.

| Label | Lattice | $Er^{3+}$, at.% | $Yb^{3+}$, at.% | T, °C | t, h |
|---|---|---|---|---|---|
| Er(0.5)Yb(<0.01) | YAG | 0.5 | <0.01 | 1750 | 6 |
| Er(0.5)Yb(5) | | | 5 | | |
| Er(0.5)Yb(10) | | | 10 | | |
| Er(0.5)Yb(20) | | | 20 | | |

X-ray diffraction (XRD), optical transmittance and luminescence measurements were used for characterization of the samples. Before measurements, the samples were ground and polished to obtain mirror-like parallel surfaces. After processing, the samples with diameter of 15 mm and 1 mm thickness were obtained. The X-ray diffraction spectra were measured using Rigaku SmartLab 3-kW X-ray powder diffractometer equipped with a copper X-ray tube operating at 40 kV and 30 mA, and a D/tex Ultra 250 solid-state detector. Phase analysis was performed on bulk samples in continuous mode in Bragg-Brentano geometry (θ/2θ scan) over an angular range of 5°-120° (2θ) with a scan step of 0.01° and speed of 1.2°/min. The absorption spectrum was measurement using Varian 5E UV-VIS-NIR spectrophotometer with a spectral bandwidth of 0.5 nm. Emission spectra were obtained at room temperature using Edinburgh Instruments FLS980 fluorescence spectrometer. Two types of detectors were used: R928P side window photomultiplier tube (Hamamatsu) for the visible part of the spectrum and R5509-72 photomultiplier tube (Hamamatsu) in nitrogen-flow cooled housing for near infrared range. InGaAs laser diode operating at 940 nm was used as an excitation source. All the measurements were taken at room temperature.

## 3. Results and discussion

Purity of the samples was confirmed using powder XRD technique. Fig. 1(a-d) shows XRD patterns of $Er^{3+}$,$Yb^{3+}$:YAG transparent ceramics. Collected XRD patterns correspond to the YAG ($Y_3Al_5O_{12}$) structure with Ia3d space group [21]. The Rietveld analysis of the obtained samples was performed using FullProf Suite software package. The calculated lattice parameters of Er(0.5)Yb(<0.01), Er(0.5)Yb(5), Er(0.5)Yb(10), and Er(0.5)Yb(20), were 12.0048(1) Å, 12.0016(1) Å, 11.9964(1) Å, and 11.9891(1) Å, respectively. The substitution of the larger $Y^{3+}$ ions (1.032 Å [22]) with the smaller $Yb^{3+}$ (0.998 Å [22]) ions caused linear decrease in the lattice parameters according to Vegard's law (Fig. 1(f)) [9]. This indicate the substitution of $Y^{3+}$ ions by $Yb^{3+}$ ions in YAG lattice.

The present paper focuses on the luminescence properties of $Er^{3+}$, and $Yb^{3+}$ ions which occupy the dodecahedral site in YAG lattice. YAG formula can be schematically write as $C_3A_2D_3O_{12}$, where C – dodecahedral, A – octahedral, D – tetrahedral, of $D_2$, $S_6$, and $S_4$ site symmetry of Schoenflies notation, respectively. Dodecahedral site have eight surrounded oxygen ions at various distance from the central cation, the first four at - $d_1$, the second four at $d_2$. The change in concentration of $Yb^{3+}$ ions caused the change of $d_1$ (2.363 Å - 2.386 Å ), and $d_2$ (2.302 Å - Å 2.274) distances, the increase and decreases, respectively (Fig. 1(e)). This caused to increase in the bond distortion of dodecahedral site from 0.017 % to 0.059 % (Fig. 1(f). The details can be found in the Table S1. It is expected to detect the change in spectroscopic properties of the both $Er^{3+}$, and $Yb^{3+}$ ions with the change in bond distortion [23]. However such changes is difficult to separate from the influence of the change in concentration of $Yb^{3+}$ ions or change in microstructure.

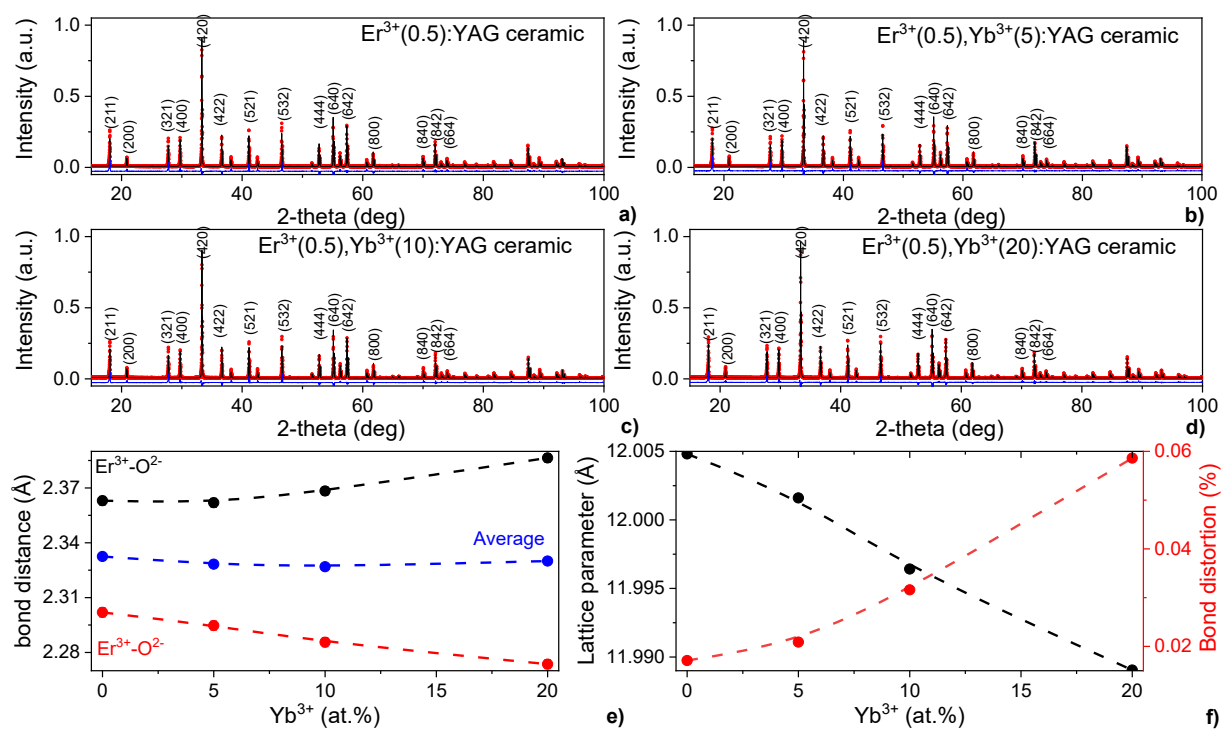

Fig. 1: X-ray diffraction pattern of the $Er^{3+}$,$Yb^{3+}$:YAG transparent ceramics (black dots), results of Rietveld refinement analysis (red curve) and their difference (blue curve) for a) Er(0.5)Yb(<0.01), b) Er(0.5)Yb(5), c) Er(0.5)Yb(10), and d) Er(0.5)Yb(20) samples; e) the change of bond lengths (black, and red dot) and their average distance (blue dot) of dodecahedral site ($Y^{3+}$, $Yb^{3+}$, $Er^{3+}$); f) the change of lattice parameter (black dots) and bond distortion of dodecahedral site (red dots) with the change on $Yb^{3+}$ concentrations.

The microstructure of $Er^{3+}$,$Yb^{3+}$:YAG transparent ceramics was investigated by Scanning Electron Microscopy (SEM). Fig. 2 shows the SEM images of the samples. No difference in the grain size distribution and the average grain sizes were found between the samples. The grain size varied from 5 μm to 25 μm with an average grain size of 18(8) μm. Er(0.5)Yb(<0.01) and Er(0.5)Yb(5) samples show similarities in the microstructure with the absence of pores or second phases. SEM images of Er(0.5)Yb(10) and Er(0.5)Yb(20) samples show the presence of a small concentration of grain boundary pores with a diameter of several microns. The sample with a higher concentration of $Yb^{3+}$ ions has higher pore concentrations at the grain boundary (Fig. S2). No pores were found in the grain volume of the samples.

Increase the concentration of $Yb^{3+}$ ions to 20 at.% caused the appearance of the impurity phases at the grain boundary. Figs. 2(e) and S5 show the SEM images and results of Energy-Dispersive x-ray Spectroscopy (EDS) analysis for Er(0.5)Yb(20) sample. Table S2 shows the quantitative results of the EDS analysis. These impurity phases have an irregular shape and size in the range of ~0.5-2.5 μm. EDS analysis of the grain volume of Er(0.5)Yb(20) sample has shown the presence of the $Al^{3+}$, $Y^{3+}$, $Yb^{3+}$, and $Er^{3+}$ ions in the YAG stochiometric ratio. No other elements were found in the matrices. EDS analysis of the inclusions shows the excess of the $Y^{3+}$ ions indicating that the impurity phases mainly consist of $Y_2O_3$ (Fig. S3). The XRD study shows an absence of the impurity phases in the samples. this indicates that their concentrations should be less than the sensitivity of X-ray analysis (0.1 wt.%). The depth of EDS analysis typically is several microns. Therefore, the collected EDS spectra contain signal not only from impurities, but also from the matrices [24]. So, the exact composition of these impurity phases remains unknown. It should be noted that the impurity phases were found in the middle of the samples within the diameters of 12 mm. The edge of ceramics was impurity-free. Fig. S4 shows the SEM images of Er(0.5)Yb(20) taken in the middle and on the edge of the sample.

EDS analysis of the impurity phases shows the presence of Si, and high concentrations of both additives, $Er^{3+}$ and $Yb^{3+}$ ions. This indicate the segregation of the sintering additive and the dopants to the impurity phase during solid-state sintering. According to the current understanding, the presence of sintering aid causes the formation of impurity phases with low melting temperature at the earlier stage of vacuum sintering. These phases decomposed to initial elements followed by their incorporation into YAG lattice [24,25]. Incorporation of the $Si^{4+}$ ions into the YAG lattice occurs through the formation of cation vacancies. This increase diffusion rates and helps to eliminate residual porosity at the final stage of the ceramic sintering [26].

$Er^{3+},Yb^{3+}$:YAG transparent ceramics were made by solid-state sintering using $Al_2O_3$, $Y_2O_3$, $Er_2O_3$, and $Yb_2O_3$ as initial components and sintering aid. Formation of garnet phase occurs through the interaction between the initial components through the YAM, and YAP intermediate phases to YAG phase in the temperature interval of 1000 – 1500°C [27,28]. Other intermediate phases can be observed in this temperature interval as well [24]. However they should disappear at the end of ceramics sintering process [28]. The presence of $Si^{4+}$ in the impurity phases indicates that the decomposition of Si-rich intermediate phases do not occur for the defined $Er^{3+},Yb^{3+}$:YAG transparent ceramics. Similar cases were detected earlier for Ce,Ca:YAG transparent ceramics, where such a behavior was explained thorough the interaction of CaO and $SiO_2$ [24]. The large concentrations of both $Er^{3+}$, and $Yb^{3+}$ ions indicates that these additives are present as well.

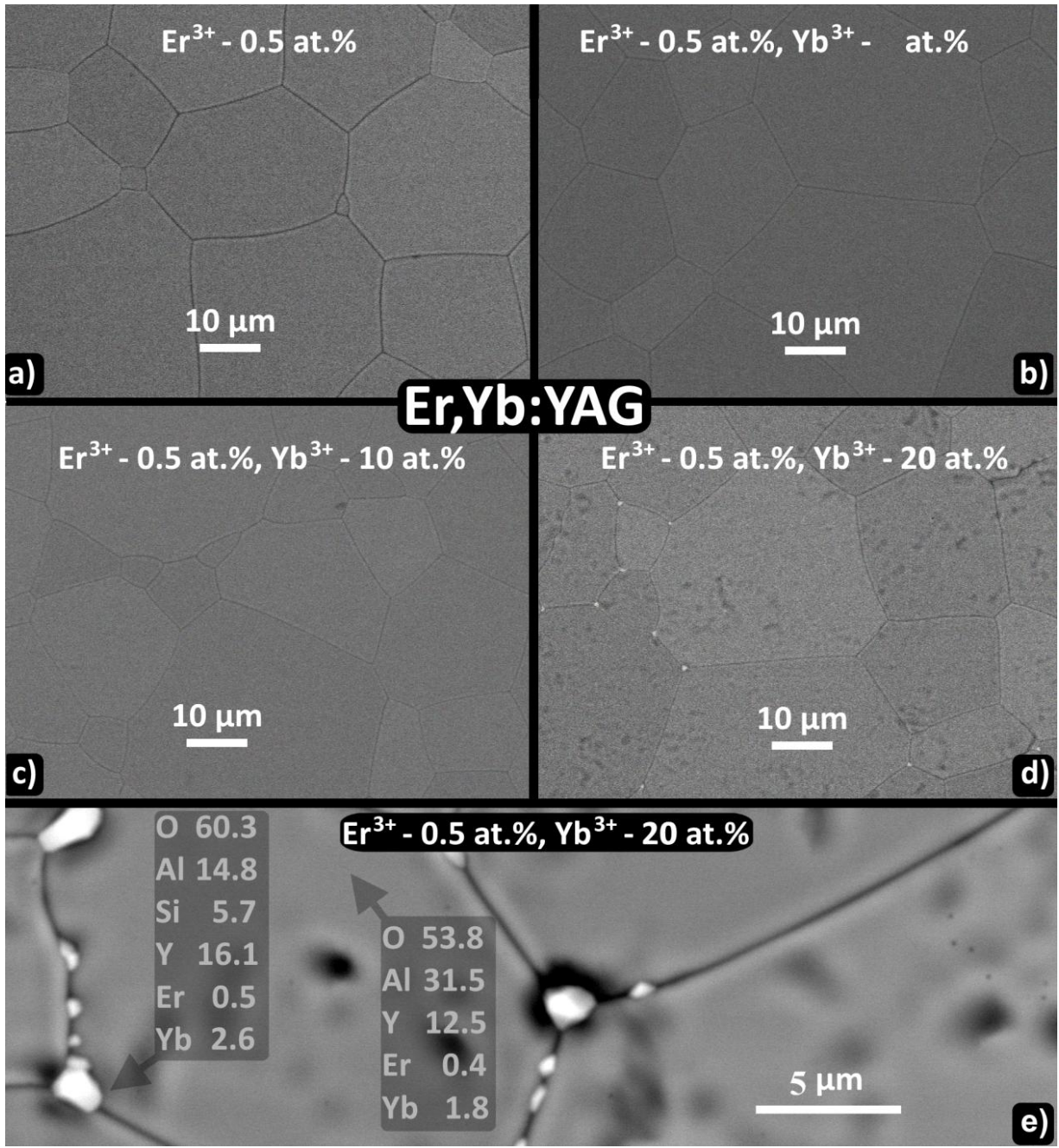

Fig. 2: SEM images of $Er^{3+},Yb^{3+}$:YAG transparent ceramics: a) Er(0.5)Yb(<0.01), b) Er(0.5)Yb(5), c) Er(0.5)Yb(10), and d), e) Er(0.5)Yb(20) samples.

The microstructure of the ceramics affects their transparency. Transmission spectra of $Er^{3+},Yb^{3+}$:YAG transparent ceramics measured in the range from 200 to 2500 nm at room temperature are shown in the Fig. 3. The increase in the concentration of $Yb^{3+}$ ions caused a decrease in the transparency of the synthesized $Er^{3+},Yb^{3+}$:YAG ceramics. In-line transmittances of Er(0.5)Yb(<0.01), Er(0.5)Yb(5), Er(0.5)Yb(10), and Er(0.5)Yb(20) samples were 81%, 81%, 78%, and 47%, respectively at 1100 nm. The samples were sintered in the same conditions in order to compare their properties. The decrease in the in-line transmittance of Er(0.5)Yb(10) sample is due to the presence of the residual porosity in the sample (Fig. S2). The further reduction in the in-line transmittance of Er(0.5)Yb(20) sample is due to the presence of both the residual porosity and the impurity phases. The sintering of YAG transparent ceramics using TEOS sintering additive requires precise tuning of $SiO_2$ concentration [26]. The presence of Si-rich impurity phases indicates that the part of silica is stack in these impurity phases and, therefore, do not participate in the ceramic sintering process. This effect reduces the positive influence of sintering additive on the sintering trajectory of $Er^{3+},Yb^{3+}$:YAG ceramics thus decreasing its transparency (Fig. 3). Therefore, the sintering of high quality $Er^{3+},Yb^{3+}$:YAG ceramics requires precise tuning TAOS concentration for each composition.

The transmission spectra contain sharp and narrow lines covering whole measured regions and corresponding to f-f transitions of $Er^{3+}$ ions. The high-resolution absorption spectra can be found in Fig. S5. No difference in the position or the height of the $Er^{3+}$ lines was found confirming the similar concentration of the $Er^{3+}$ ions in the synthesized $Er^{3+},Yb^{3+}$:YAG transparent ceramics. The most intense lines of the $Er^{3+}$ ions are observed at 256.0 nm, 380.9 nm, 488.0 nm, 524.3 nm, 647.3 nm, 966.5 nm, 1475.3 nm, and 1532.5 nm. Absorption in the range 850 - 1000 nm correspond to $^2F_{7/2} \rightarrow {}^2F_{5/2}$ $Yb^{3+}$ transitions. The main lines of $Yb^{3+}$ ions are observed at 915 nm, 927 nm, 939 nm, 969 nm, and 1030 nm.

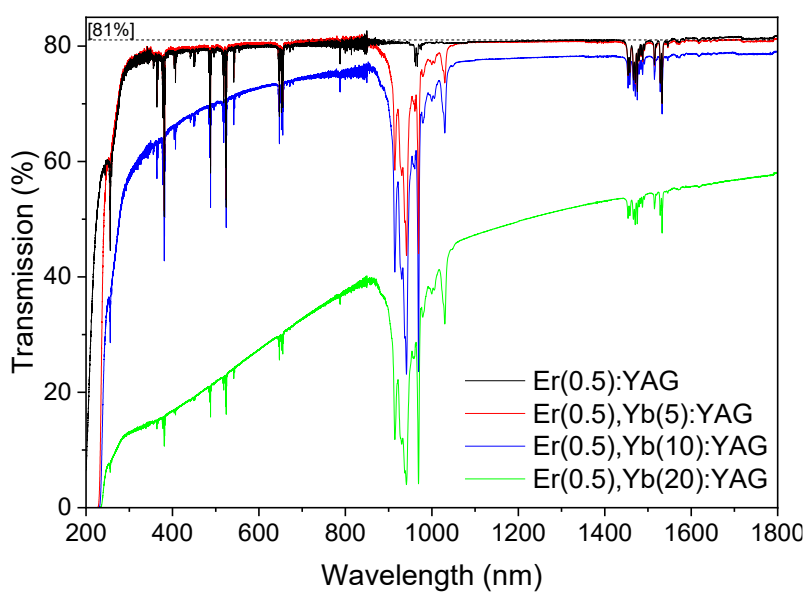

Fig. 3: Optical absorption spectra of $Er^{3+},Yb^{3+}$:YAG transparent ceramics.

The special attention should be paid to the Er(0.5)Yb(20) sample with two different regions i.e. middle part with impurities and the edge. Fig. S6 shows absorption spectra of Er(0.5)Yb(20) measured at the edge and in the middle of the sample. No difference in the intensity and position of the liens was found indicating the same concentrations of both $Er^{3+}$ and $Yb^{3+}$ ions at the edge and in the middle of Er(0.5)Yb(20) sample. At the same time the emission spectra show a difference in the shape of $Yb^{3+}$ band measured at the edge and in the middle of the sample (Fig. 4(a)).

Fig. 4(a) shows emission spectra of Er(0.5)Yb(20) ceramics taken at the edge and in the middle of the sample. The emission spectra measured from the edge show the shape similar to the one observed for other samples and Yb:YAG single crystal [29], while the emission spectra of $Yb^{3+}$ ions measured from the edge is broader (Fig. 4(a)). The most intense emission lines of $Yb^{3+}$ ions were observed at 1028.7 nm and 1031.5 nm for the edge and the middle part of the sample, respectively. The emission spectrum taken in the middle of Er(0.5)Yb(20) sample is probably the superposition of the emission spectra of $Yb^{3+}$ ions in YAG lattice and impurities (most likely $Yb^{3+}$:$Y_2O_3$ [6]. Other possible explanation is the agglomeration of $Yb^{3+}$ ions in YAG lattice (Fig. 2(c)). We suppose that the results obtained for Er(0.5)Yb(20) sample are determined by optimal concentration of TAOS sintering additive, however, detailed explanation requires further investigations. For comparison of the spectroscopic properties of $Er^{3+}$,$Yb^{3+}$:YAG transparent ceramics, the emission spectra of Er(0.5)Yb(20) sample were measured for the part without impurity phases (the edge).

Luminescence spectra of $Er^{3+}$,$Yb^{3+}$:YAG transparent ceramics were obtained at 940 nm excitation (Fig. 4). These spectra were normalized to the most intense $Er^{3+}$ emission line for each Stokes and anti-Stokes part (Fig. 4(c,d)). The Stokes part of the spectra consists of the emission lines of $Yb^{3+}$ and $Er^{3+}$ ions with the most intense lines at 1028.7 nm and 1530.8 nm, respectively (Fig. 4(c)). Emission of $Yb^{3+}$ ions originates from $^2F_{5/2} \rightarrow {}^2F_{7/2}$ electron transition [30,31]. Normalized emission spectra of $Yb^{3+}$ ions are shown in the Fig. 4(b). It should be noted that $Yb^{3+}$ emission was observed in Er(0.5)Yb(<0.01) sample indicating the presence of small concentrations of $Yb^{3+}$ ions (as impurity with $Y^{3+}$ ions). The $Yb^{3+}$ concentration has no influence on the position of emission maximum. The most intense emission lines are observed at 967.1 nm, 1005 nm, 1028.7 nm, and 1046.8 nm [29,32]. However, the increase in the $Yb^{3+}$ concentration caused a decrease in the high energy tail of the emission spectra (967.1 nm) and at the same time increase in the low energy tail of emission spectra (1046.8 nm). When the emission spectra are normalized, the intensities of 967.1 nm line were 0.41, 0.28, 0.18 and 0.1 for Er(0.5)Yb(<0.01), Er(0.5)Yb(5), Er(0.5)Yb(10), and Er(0.5)Yb(20) samples, respectively. The emission intensities of 1046.8 nm line were 0.1, 0.16, 0.22, 0.31 for Er(0.5)Yb(<0.01), Er(0.5)Yb(5), Er(0.5)Yb(10), and Er(0.5)Yb(20) samples, respectively.

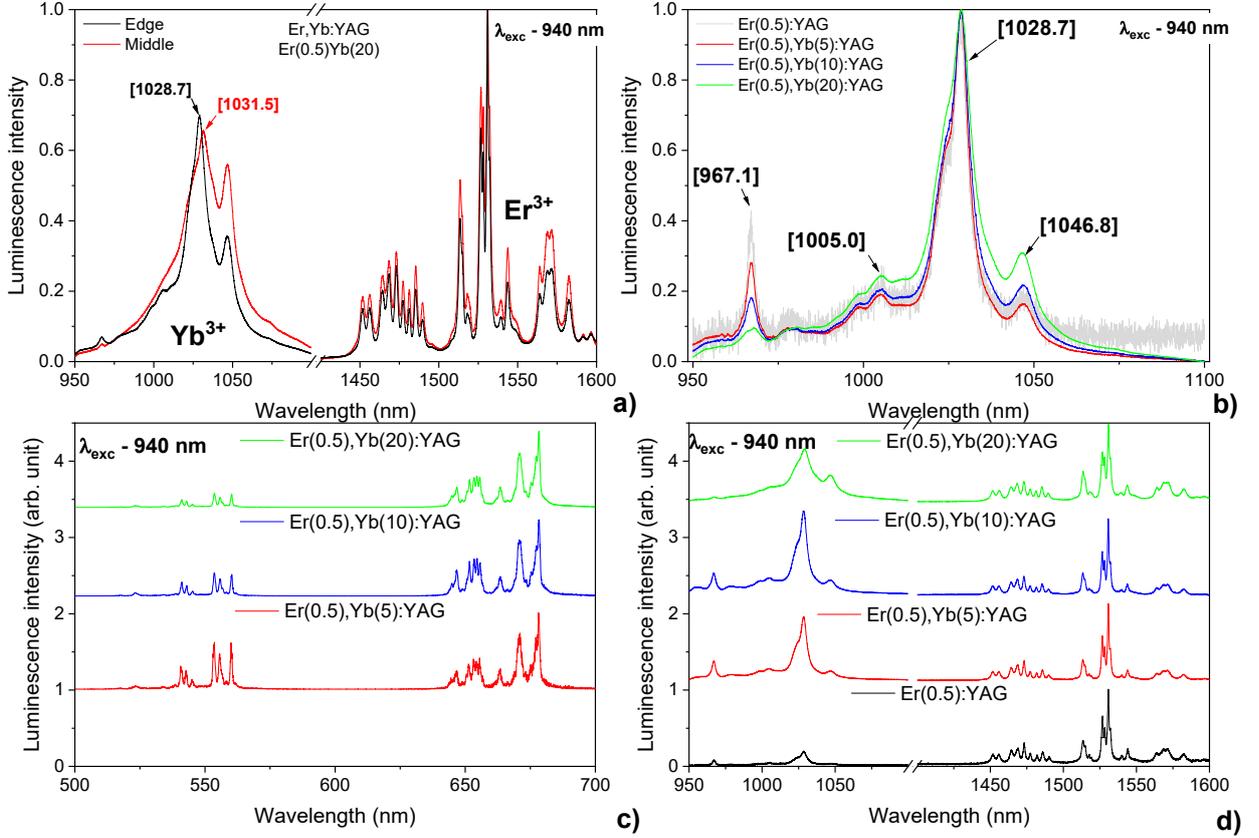

Fig. 4: Emission spectra of $Er^{3+},Yb^{3+}$:YAG transparent ceramics ($\lambda_{exc}$ = 940 nm, T – 300 K): a) Er(0.5)Yb(20) sample measured at the edge (black line) and in the middle (red line); b) normalized emission spectra of $Yb^{3+}$ ions; c) anti-Stokes and d) stokes emission.

We suppose that the change in the shape of emission spectra is due to an increase in the neat transfer between $Er^{3+}$ and $Yb^{3+}$ ions. Due to similar energy differences for $^4I_{11/2} \rightarrow ^4I_{15/2}$ and $^2F_{7/2} \rightarrow ^2F_{5/2}$ electron transitions, energy transfer could occur in both directions including direct $Yb^{3+} \rightarrow Er^{3+}$ and back $Er^{3+} \rightarrow Yb^{3+}$ transfer [33]. The possible explanation of the difference in the emission spectra could be connected with the electronic structures of $Er^{3+}$ and $Yb^{3+}$ ions. The energy scheme of $Yb^{3+}$ ion contains two Stark multiplets, $^2F_{5/2}$ and $^2F_{7/2}$ with energy difference of ~10 000 cm$^{-1}$ (Fig. 5). In a crystal field of cubic symmetry the six-fold degenerated J = 5/2 manifold is split into Γ7 doublet (2') and Γ8 quartet (0', 1'), and for octahedral (six-fold) coordination the quartet has lower energy. The J = 7/2 manifold forms Γ8 quartet (1, 2) and two doublets (Γ6 (0) and Γ 7 (3)) [6]. The energy scheme of $Er^{3+}$ ions is more complicated, and the most important for this work are three lowest energy levels ($^4I_{15/2}$, $^4I_{13/2}$, and $^4I_{11/2}$). Each of them split into bunch of Stark components with energy difference of less than 50 cm$^{-1}$ providing rich absorption and emission spectra.

Proposed energy scheme is shown in Fig. 5. The positions of energy levels and Stark components of Yb$^{3+}$ were estimated based on the absorption and emission spectra of Er$^{3+}$,Yb$^{3+}$:YAG transparent ceramics, as described in our earlier paper [6]. The positions of energy levels and Stark components of Er$^{3+}$ ion were taken from literature [34], and well agree with absorption and emission spectra of Er$^{3+}$,Yb$^{3+}$:YAG transparent ceramics. The photon energy of 940 nm laser diode corresponds to the energy difference between the lowest Stark components (0) of the $^2F_{7/2}$ energy level and the second lowest Stark components (1') of the $^2F_{5/2}$ energy level. It should be noted that the energy level diagram presented in Fig. 5 is valid for ions located in a high-symmetry site. In our case, distortion of the dodecahedral site caused additional splitting. It is assumed that this additional splitting has only a minor effect on the luminescence spectra.

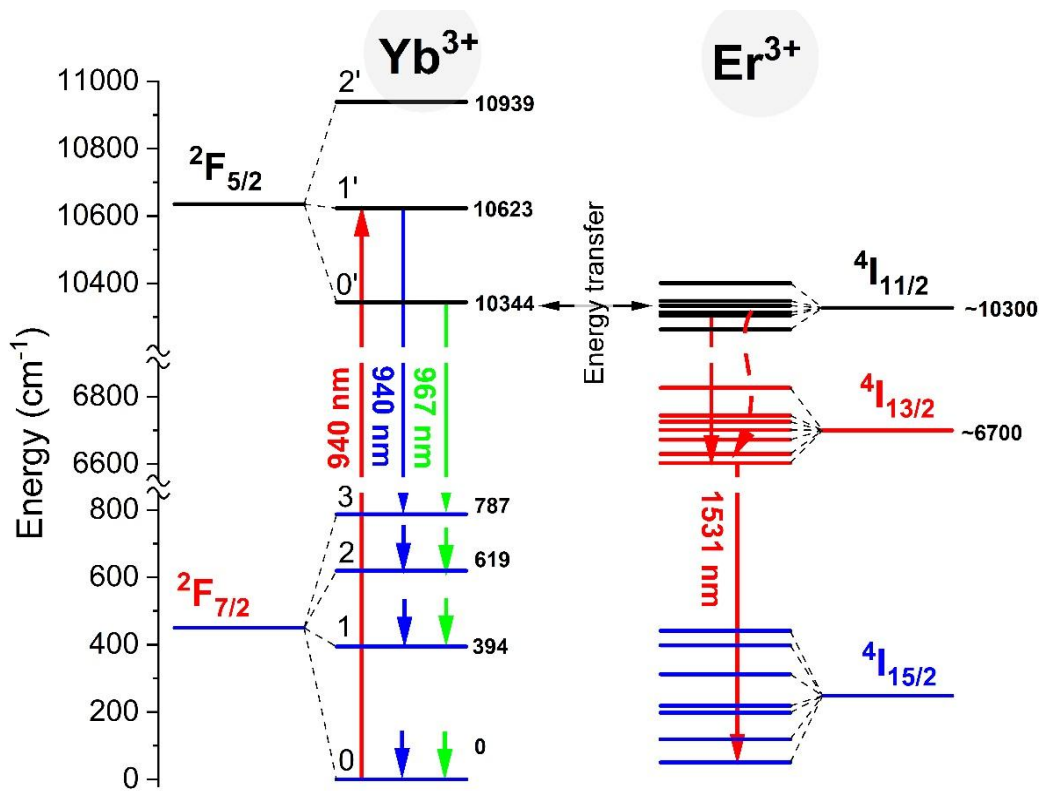

Fig. 5: Schematic illustration of the energy levels and Stark components of Er$^{3+}$ and Yb$^{3+}$ ions in Er$^{3+}$,Yb$^{3+}$:YAG transparent ceramics.

In the case of pumping at $\lambda_{exc}$ = 940 nm, a considerable part of $Yb^{3+}$ ions are found in the upper manifold. Fast intraband thermalization leads to distribution of electrons between Stark manifolds with most of the excited ions being in (0´). Further, these electrons relax to the ground state with energy transfer to $Er^{3+}$ ions or produce photon. Both of these processes caused electron from 0´ Stark component of $^2F_{5/2}$ energy level to one of the Stark components of $^2F_{7/2}$ ground state. Due to the better match in energy difference, transitions between the Stark component with higher energy differences, such as 0´→0 or 0´→1 have higher probability to take part in energy transfer ($Yb^{3+}$→$Er^{3+}$) than the transitions with smaller difference such as 0´→2 or 0´→3. The increase in the $Yb^{3+}$ concentration causes an increase in the $Yb^{3+}$→$Er^{3+}$ energy transfer probability, thus depleting the high energy components. The other possible explanation can be related to the radiation trapping effect. The mechanism is similar, but involves radiative energy transfer between $Yb^{3+}$ ions. Detailed explanation can be found in our earlier paper [6]. Most probably both mechanisms are responsible for the change in the emission spectra of $Yb^{3+}$ ions. It should be noted that the increase in the bond distortion (Fig. 1(f)) could be the reason of the changes in luminescence properties of $Yb^{3+}$ ions.

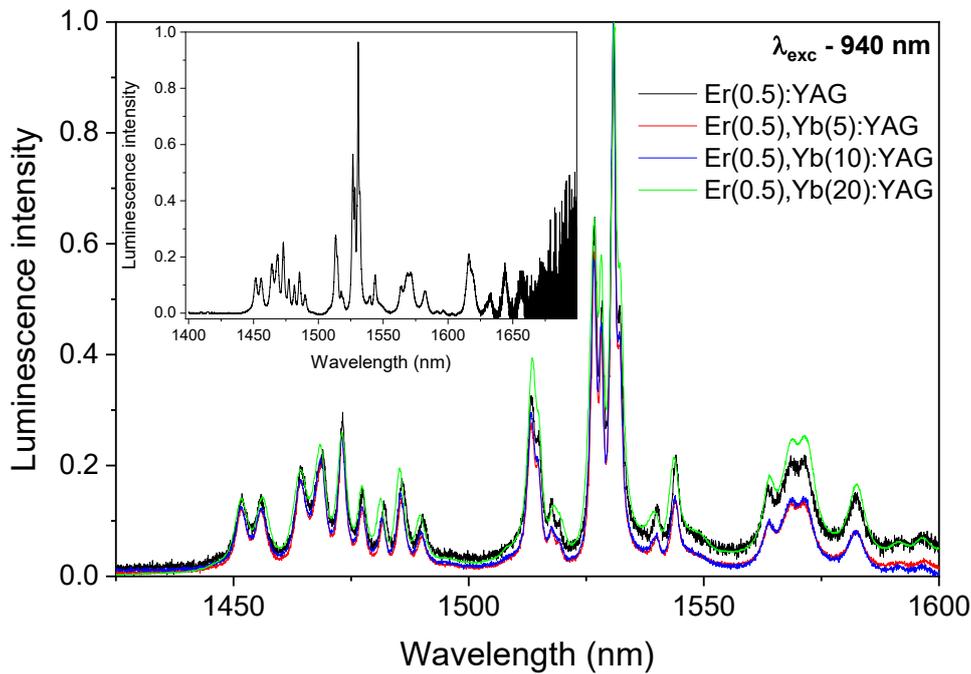

Fig. 7b: Normalized emission spectra of $Er^{3+}$ ions measured in the range of 1400–1600 nm in $Er^{3+}$,$Yb^{3+}$:YAG transparent ceramics. The inset shows the emission spectrum of Er(0.5)Yb(20) sample measured in the range of 1400-1700 nm with a manually corrected background.

The change in the concentration of $Yb^{3+}$ ions had a minor influence on the emission shape of $Er^{3+}$ ions. Irradiation of the samples by 940 nm light caused an appearance of $Er^{3+}$ emission with the most intense line at 1530.8 nm originated from $^4I_{13/2} \rightarrow {}^4I_{15/2}$ transition (Fig 4(d)). Fig. 7b shows normalized emission spectra of $Er^{3+}$ ions measured in the range of 1400 nm - 1600 nm. The most intense emission lines were observed at 1473.1 nm, 1513.4 nm, 1530.8 nm, and 1571.4 nm. The minor changes in the positions of the emission lines were found for Er(0.5)Yb(20) samples, where emission maxima differ by ~ 0.2 nm in comparison to the other samples that can be explained by the influence of the $Yb^{3+}$ ions in the nearby sites. The increase in the $Yb^{3+}$ concentration increase the possibility to find $Yb^{3+}$ ions in the nearby sites of $Er^{3+}$ ions. Due to the difference in the ionic radii of $Yb^{3+}$ and $Y^{3+}$ ions, the presence of $Yb^{3+}$ ions in nearby dodecahedral sites changes the $Er^{3+}$ - $O^{2-}$ ions distance (Fig. 1(e)) modifying their interaction. It should be noted that the full $Er^{3+}$ emission spectrum extend up to 1670 nm. However, the detector sensitivity and background correction ability are not sufficient to collect high quality data. To prevent misunderstanding, this paper present emission spectra only up to 1600 nm. The full-range emission spectrum with a manually corrected background is shown in the inset of Fig. 7b.

The change in the concentration of $Yb^{3+}$ ions changes $Er^{3+}/Yb^{3+}$ emission intensity ratio, which was 4, 0.7, 0.5, and 0.85 for Er(0.5)Yb(<0.01), Er(0.5)Yb(5), Er(0.5)Yb(10), and Er(0.5)Yb(20) samples, respectively. No pattern was found for $Er^{3+}/Yb^{3+}$ emission ratio in the ceramics with different concentration of $Yb^{3+}$ ions. It should be noted that the emission of Er(0.5)Yb(<0.01) sample was recorded at the excitation power of two orders of magnitude larger than for the other ones. $Er^{3+}/Yb^{3+}$ emission ratio is determined by population of $^4I_{11/2}$ and $^2F_{5/2}$ levels of $Er^{3+}$ and $Yb^{3+}$ ions, which depends on $Er^{3+} \leftrightarrow Yb^{3+}$ energy back transfer, up-conversion, and quenching mechanisms [35].

Irradiation of the sample by 940 nm light caused an appearance of green emission originated from $Er^{3+}$ ions due to upconversion process that occurs in $Er^{3+}$-$Yb^{3+}$ ion pair [36]. Fig. 4(c) shows the visible part of the emission spectra of $Er^{3+},Yb^{3+}$:YAG transparent ceramics at $\lambda_{exc}$ = 940 nm. The visible emission spectra of $Er^{3+}$ ions can be divided into two parts: red emission with the most intense lines at ~857 nm, ~842 nm, ~828 nm, ~809 nm, ~794 nm, and green emission with the most intense lines at ~622 nm, ~613 nm, ~608 nm, ~587 and ~583 nm. This emission originated from $^2H_{11/2}, {}^4S_{3/2} \rightarrow {}^4I_{15/2}$ and $^4F_{9/2} \rightarrow {}^4I_{15/2}$ transitions corresponds to ~ 520-560 nm green and 640-680 nm red emission, respectively. Due to low emission intensity, the visible part of the emission spectra was recorded with the ~0.5 nm step. Within this accuracy, No difference in the position of the emission maximums was observed at such resolution.

The intensity ratio of the $^2H_{11/2} \rightarrow {}^4I_{15/2}$ and $^4S_{3/2} \rightarrow {}^4I_{15/2}$ emission bands of $Er^{3+}$ ions is temperature-dependent, enabling the estimation of the temperature of the irradiated sample (temperature sensor) [37]. The population of the $^2H_{11/2}$ and $^4S_{3/2}$ states follows the Maxwell-Boltzmann distribution, as described by the formula:

$$\frac{I_{2H11/2 \rightarrow 4I15/2}}{I_{4S3/2 \rightarrow 4I15/2}} = A \times \exp\left(\frac{\Delta E}{kT}\right)$$

where $I_{2H11/2 \rightarrow 4I15/2}$ and $I_{4S3/2 \rightarrow 4I15/2}$ are the populations of energy levels, A is a constant dependent on degeneracies of these levels, $\Delta E$ is energy between levels, T is temperature in Kelvins and k is the Boltzmann constant.

The $^2H_{11/2} \rightarrow {}^4I_{15/2}$ to $^4S_{3/2} \rightarrow {}^4I_{15/2}$ intensity ratio were 0.463, 0.496, and 0.525 for Er(0.5)Yb(5), Er(0.5)Yb(10), and Er(0.5)Yb(20) samples. The results of the integral intensity ratio for the $^2H_{11/2} \rightarrow {}^4I_{15/2}$ and $^4S_{3/2} \rightarrow {}^4I_{15/2}$ transitions are shown in Fig. S7. The increase in the value of the ratio shows that, as the concentration of $Yb^{3+}$ ions increases with excitation under the same conditions, the local temperature around $Er^{3+}$ ions increases. As was discussed in [38] increasing the sample temperature affects the ratio of green to red emission intensity. The calculated green to red emission ratio were 0.46, 0.21, and 0.13 for Er(0.5)Yb(5), Er(0.5)Yb(10), and Er(0.5)Yb(20) samples. The other explanation to such change can be caused by the deformation of the dodecahedral site (Fig. 1(f)). The changes in $Er^{3+} - O^{2-}$ bond distortion affects the green to red emission ration of $Er^{3+}$ ions [39]. This shows that by the amount of ytterbium ions, the colour of the upconversion emission can be adjusted.

The energy transfer processes between $Er^{3+}$ and $Yb^{3+}$ ions were investigated by measuring the lifetimes of 1531 nm and 1028 nm luminescence. The luminescence decay curves of the $Er^{3+},Yb^{3+}$:YAG transparent ceramics at $\lambda_{exc}$ = 940 nm are shown in the Fig. 6. The shapes of the decay curves of $Yb^{3+}$ ions can be described by single exponential function using the following equation:

$$I = I_0 + A_1 e^{(-t/\tau)}$$

where I is the luminescence intensity at time t, $\tau$ is lifetime, $A_1$ is constant. The lifetimes of $Yb^{3+}$ ions increase with $Yb^{3+}$ concentration. The measured lifetimes of Er(0.5)Yb(<0.01), Er(0.5)Yb(5), Er(0.5)Yb(10), and Er(0.5)Yb(20) samples at $\lambda_{em}$ = 1028 nm are 0.76 ms, 0.74 ms, 0.91 ms, and 1.16 ms, respectively. The calculated lifetimes are collected in Table 2. The increase in luminescence lifetimes of $Yb^{3+}$ ions with their concentration in YAG lattice is a common trend for transparent ceramics. The lifetimes of Yb:YAG ceramics increased from 1.0 ms (1 at.%) to 1.4 ms (15 at.% of $Yb^{3+}$ ions) [40].

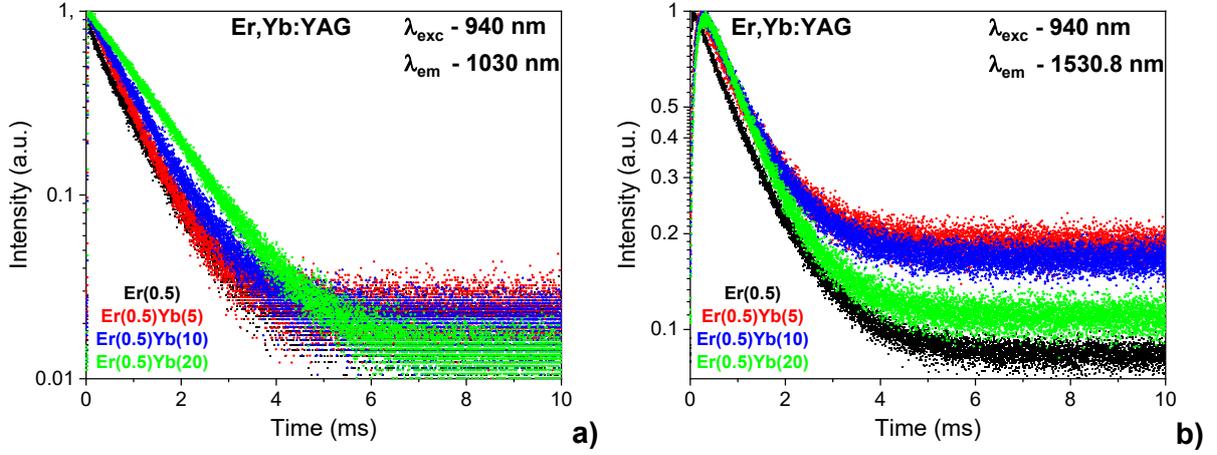

Fig. 6: Luminescence decay curves of 1531 nm and 1028 nm emission of transparent $Er^{3+},Yb^{3+}$:YAG ceramics at $\lambda_{exc}$ = 940 nm.

The luminescence decay curves of $Er^{3+}$ luminescence were calculated using the special exponential function. This function can describe the whole curve including the initial growth part using two parameters: $\tau_g$ and $\tau_d$.

$$I = \begin{cases} I_0 + A_d + A_g\left(e^{-t_c/\tau_g} - e^{-t/\tau_g}\right) & t \leq t_c \\ I_0 + A_d e^{-(t-t_c)/\tau_d} & t > t_c \end{cases},$$

where I is the luminescence intensity at time t, $\tau_g$ and $\tau_d$ are growth and decay components, respectively, $A_g$ and $A_d$ are constants, $t_c$ is critical time when emission growth change to emission decay. The concentration of $Yb^{3+}$ ions has insufficient influence on the luminescence decay times of $Er^{3+}$ ions, with the exception of Er(0.5)Yb(<0.01) samples. The measured $\tau_d$ of Er(0.5)Yb(<0.01), Er(0.5)Yb(5), Er(0.5)Yb(10), and Er(0.5)Yb(20) samples at $\lambda_{em}$ = 1531 nm are 1.01 ms, 0.95 ms, 0.95 ms, and 0.93 ms, respectively. The increase in the concentration of $Yb^{3+}$ ions caused an increase in the recorded growth lifetimes of $Er^{3+}$ ions. The calculated $\tau_d$ of Er(0.5)Yb(<0.01), Er(0.5)Yb(5), Er(0.5)Yb(10), and Er(0.5)Yb(20) samples at $\lambda_{em}$ = 1531 nm were 0.011, 0.062, 0.072, and 0.083, respectively. The calculated lifetimes are collected in Table 2.

Table 2: Measured lifetimes of $Yb^{3+}$ and $Er^{3+}$ ions in $Er^{3+}$,$Yb^{3+}$:YAG transparent ceramics: lifetimes of $Yb^{3+}$ ($\tau_1$) at $\lambda_{exc}$ = 940 nm and $\lambda_{em}$ = 1028 nm, $Er^{3+}$ ions luminescence growth ($\tau_g$), and decay ($\tau_d$) times at $\lambda_{exc}$ = 940 nm and $\lambda_{em}$ = 1531 nm.

| Label | Yb, at.% | Er, at.% | $Yb^{3+}$ ($\tau_1$), ms | $Er^{3+}$ ($\tau_d$) ($^4I_{13/2}$), ms | $Er^{3+}$ ($\tau_g$) ($^4I_{11/2}$), ms |
|---|---|---|---|---|---|
| Er(0.5) | <0.01 | 0.5 | 0.755(1) | 1.013(1) | 0.011(1) |
| Er(0.5)Yb(5) | 5 | 0.5 | 0.738(1) | 0.946(2) | 0.062(1) |
| Er(0.5)Yb(10) | 10 | 0.5 | 0.907(1) | 0.953(2) | 0.072(1) |
| Er(0.5)Yb(20) | 20 | 0.5 | 1.175(1) | 0.930(2) | 0.083(1) |

The spectroscopic properties of $Er^{3+}$,$Yb^{3+}$:YAG transparent ceramics are based on the energy transfer between $Yb^{3+}$ and $Er^{3+}$ ions which act as donor and acceptor, respectively. Due to close values of $^2F_{5/2}$ and $^4I_{11/2}$ energy levels of $Yb^{3+}$ and $Er^{3+}$ ions, respectively, both direct energy transfer and back transfer can occur [33]. The energy transfer between erbium ions complicates the relaxation processes in $Er^{3+}$,$Yb^{3+}$:YAG transparent ceramics. Therefore, the common approach to describe the energy transfer is not suitable for our case.

First of all, let us consider the changes in the radiative recombination rates of $Er^{3+}$ ions at 1531 nm. Irradiation of the $Er^{3+}$,$Yb^{3+}$:YAG transparent ceramics by 940 nm light leads to $^2F_{7/2} \rightarrow ^2F_{5/2}$ transition of $Yb^{3+}$ ions with following radiative recombination ($Yb^{3+}$ emission) or energy transfer to $^4I_{11/2}$ electronic state of $Er^{3+}$ ions. The increase in the measured lifetime of $Yb^{3+}$ ions can be explained by the increase in the probability of $Er^{3+} \rightarrow Yb^{3+}$ energy back transfer. It should be noted that if $Er^{3+}$ ions act as acceptors, the energy migration between ytterbium ions can be neglected [33].

The energy transfer process between $Yb^{3+}$ and $Er^{3+}$ ions occurs thorough the population of $^4I_{11/2}$ energy level of $Er^{3+}$ ions. Depopulation of $^4I_{11/2}$ energy level can occur in the several ways: 1) energy back transfer process to $Yb^{3+}$ ions; 2) $^4I_{11/2} \rightarrow ^4I_{15/2}$ radiative transition; 3) $^4I_{11/2} \rightarrow ^4I_{13/2}$ radiative or nonradiative transition; 4) energy transfer to structural defects. The measured 1531 nm emission originated from $^4I_{13/2} \rightarrow ^4I_{15/2}$ electron transition and consists of growth ($\tau_g$) and decay ($\tau_d$) parts. The growth is part appears in the luminescence decay curves due to the population of $^4I_{13/2}$ emitting level from $^4I_{11/2}$ higher energy states. Thus it might be concluded that the measured values of $\tau_g$ and $\tau_d$ at $\lambda_{em}$ = 1531 nm correspond to the lifetimes of $^4I_{11/2}$, and $^4I_{13/2}$ levels, respectively. The similarity in the measured values of $\tau_d$ (Table 2) for the samples indicates the weak energy transfer process from $Er^{3+}$ ions to luminescence quenching centers.

However, it is noteworthy that the $\tau_g$ growth time increases with increasing ytterbium concentration. This relationship indicates that this time is determined by the lifetime of the $^2F_{5/2}$ ytterbium ion, which feeds the $^4I_{11/2}$ $Er^{3+}$ level, and the extension of this time is related to the effect of radiation trapping and energy diffusion on $Yb^{3+}$ ions related to decreasing distance between ytterbium ions, which increases the probability of energy transfer between them [41,42].

One of the important question that should be addressed is the optimal concentration of $Er^{3+}$ and $Yb^{3+}$ additives needed to achieve the best laser performance. The laser performance of $Er^{3+},Yb^{3+}$:YAG transparent ceramics depends on multiple parameters, the most important of which is residual losses. Residual losses can be caused by absorption centers, light-scattering centers, energy transfer processes, etc.. Even a slight increase in residual losses drastically reduce laser efficiency. For example, an increase in residual porosity from ~ 0.002% to ~0.09% reduces laser efficiency from ~34% to ~2% [43].

The transparency of the synthesized samples decreases with increasing $Yb^{3+}$ concentration. Therefore, samples with the lowest concentration of $Yb^{3+}$ ions are more suitable for use as a laser host. However, ceramics transparency can be improved by adjusting the sintering parameters for a specific ceramic composition [16]. The optimal concentration of both $Er^{3+}$ and $Yb^{3+}$ ions should be determined based on their spectral characteristic.

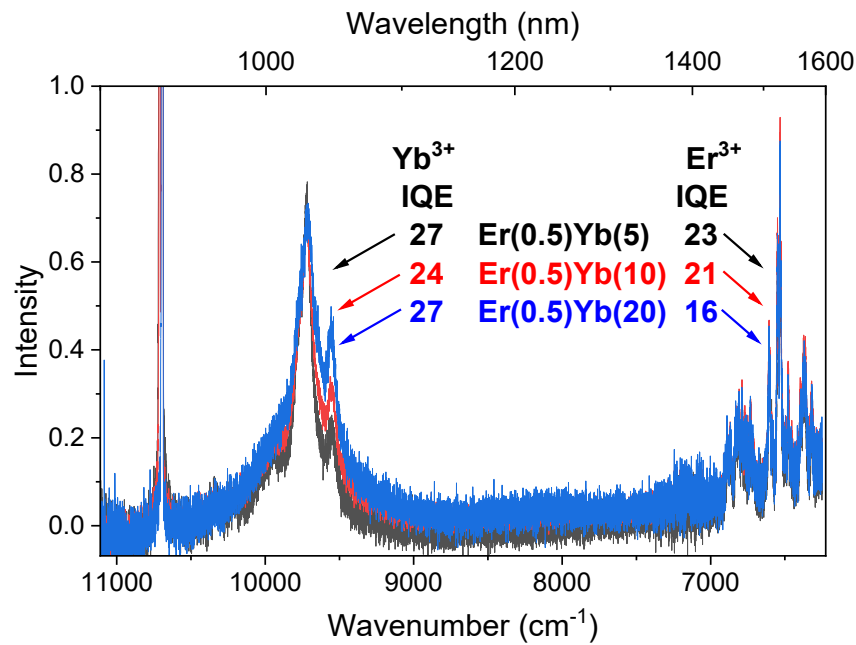

Fig. 8: The measured spectra of integration sphere setup under laser diode excitation ($\lambda_{max}$ = 940 nm, T – 295K) and $Er^{3+},Yb^{3+}$:YAG transparent ceramics inside.

Another possible mechanism for the decrease in laser performance is related to energy transfer efficiency in the samples. One of the main disadvantages of $Er^{3+}$ ions is their low absorption, which makes it difficult to use $Er^{3+}$-doped laser hosts. To overcome, $Yb^{3+}$ co-doping is used, as $Yb^{3+}$ can efficiently absorb excitation radiation and transfer energy to $Er^{3+}$ ions. However, the absorbed energy can also be transferred to structural defects, creating parasitic energy transfer channels. The efficiency of these processes can be estimated by measuring the emission spectra in an integrating sphere. Fig. 8 shows the measured spectra obtained using integration sphere setup under laser diode excitation with $Er^{3+},Yb^{3+}$:YAG transparent ceramics ($\lambda_{exc}$ - 940 nm, T - 295K). These spectra were used to calculate the internal quantum efficiency (IQE) of $Er^{3+},Yb^{3+}$:YAG ceramics luminescence in the range of 900-1600 mn (with the correction to missing part). The detailed $Yb^{3+}$ and $Er^{3+}$ emission spectra from the integration sphere setup can be found in Fig. S8.

The calculated overall IQE for Er(0.5)Yb(5), Er(0.5)Yb(10), and Er(0.5)Yb(20) samples was 50%, 45% and 43 %, respectively. The calculated IQE of $Er^{3+}$ ions was 23%, 21% and 16%, for Er(0.5)Yb(5), Er(0.5)Yb(10), and Er(0.5)Yb(20) samples, respectively. an increase in the $Yb^{3+}$ concentration from 5% to 20% led to a decrease in IQE of $Er^{3+}$ ions from 23% to 16%. Based on these results, it can be predicted that the optimal $Er^{3+}/Yb^{3+}$ ratio should be equal or less than 1/10 for $Er^{3+},Yb^{3+}$:YAG ceramics doped with no more than 0.5 at.% of $Er^{3+}$ ions. However, the proposed optimal $Er^{3+}/Yb^{3+}$ ratio is more of prediction than a define conclusion. The presented data come from samples doped the same $Er^{3+}$ concentration. It is expected that the optimal $Er^{3+}/Yb^{3+}$ ratio will decrease as the $Er^{3+}$ concentration increases. Additionally, changes in sintering condition are likely to influence the spectroscopic properties of the ceramics. Therefore, further work should include measurements of the spectroscopic properties of $Er^{3+},Yb^{3+}$:YAG ceramics with different $Er^{3+}$ concentration and/or ceramics sintered under varying conditions.

## 4. Conclusions

Concentration series of $Er^{3+},Yb^{3+}$:YAG transparent ceramics were synthesized by solid-state sintering in vacuum. The concentration of the $Er^{3+}$ ions were the same for all the samples and equal to 0.5 at.%, while the concentration of $Yb^{3+}$ ions was 0 at.%, 5 at.%, 10 at.%, and 20 at.%. In-line transmission spectra of $Er^{3+},Yb^{3+}$:YAG transparent ceramics have shown the decrease in transparency with increase of $Yb^{3+}$ concentration. In-line transmittance of the samples doped with 0 at.%, 5 at.%, 10 at.%, and 20 at.% of $Yb^{3+}$ ions was 81%, 81%, 78%, and 47% at 1100 nm, respectively. XRD analysis revealed YAG structure without impurity phases.

SEM analysis confirmed the absence of the impurity phases with the exception of the $Er^{3+},Yb^{3+}$:YAG transparent ceramics doped with 20 at.% of $Yb^{3+}$ ions where $Yb^{3+}$-rich secondary phase was found. EDS analysis of the impurity phases show the presence of Si, and larger concentrations of both additives, $Er^{3+}$, and $Yb^{3+}$ ions indicating the segregation of the sintering additive and the dopants to the impurity phase during solid-state sintering. The presence of Si-rich impurity phases indicates that the fraction of silica is bonded in these impurity phases and therefore do not participate in the ceramic sintering process reducing the positive influence of sintering aid on the sintering trajectory of $Er^{3+},Yb^{3+}$:YAG ceramics and decreasing its transparency. Therefore, the sintering of high quality $Er^{3+},Yb^{3+}$:YAG ceramics requires precise tuning of sintering aid for each composition separately.

Luminescence spectra of $Er^{3+},Yb^{3+}$:YAG transparent ceramics were measured at $\lambda_{exc}$= 940 nm. The concentration of $Yb^{3+}$ ions has no influence on the positions of $Er^{3+}$, and $Yb^{3+}$ emission maxima. However, an increase in $Yb^{3+}$ concentration decreases the intensity of high energy emission with an increase in the low energy part of $Yb^{3+}$ emission. The possible explanation of this effect is an increase of neat transfer between $Er^{3+}$ and $Yb^{3+}$ or radiative energy transfer between erbium ions. The measured lifetimes of $Yb^{3+}$ ions in $Er^{3+},Yb^{3+}$:YAG transparent ceramics increased with $Yb^{3+}$ concentration from 0.76 ms to 1.16 ms. The measured lifetime of $^4I_{13/2}$ energy level of $Er^{3+}$ ions was 1 ms for all samples. The calculated lifetime of $^4I_{11/2}$ energy level of $Er^{3+}$ ions increases with concentration of $Yb^{3+}$ ions. This result was explained by an increase in the efficiency of energy transfer between $Yb^{3+}$ and $Er^{3+}$ ions.

The calculated overall internal quantum efficiency for Er(0.5)Yb(5), Er(0.5)Yb(10), and Er(0.5)Yb(20) samples was 50%, 45% and 43 %, respectively. The calculated IQE of $Er^{3+}$ ions were 23%, 21% and 16%, for Er(0.5)Yb(5), Er(0.5)Yb(10), and Er(0.5)Yb(20) samples, respectively. The optimal $Er^{3+}/Yb^{3+}$ ratio should be equal or less than 1/10 for $Er^{3+},Yb^{3+}$:YAG ceramics doped with no more than 0.5 at.% of $Er^{3+}$ ions. As the $Er^{3+}$ concentration increase beyond 0.5 at.%, the optimal $Er^{3+}/Yb^{3+}$ ratio is expected to decrease.

**Acknowledgments**

A.S. This research was funded by the National Science Centre, Poland, grant number: MINIATURA-6 2022/06/X/ST7/01706. M.C: This work was supported by the Polish National Science Center, grant: OPUS 23, UMO-2022/45/B/ST5/01487. M.C, and R.T: This work has been co-financed by the European Union under the HORIZON.1.2 – Marie Skłodowska-Curie Actions (MSCA), topic HORIZON-MSCA-2023-SE-01 – MSCA Staff Exchanges 2023, Project number 101182995.